\documentclass[prx,twocolumn,superscriptaddress,longbibliography]{revtex4-2}
\usepackage{bm}
\usepackage{amsmath, amsfonts}
\usepackage{amssymb}
\usepackage{times}
\usepackage{graphicx}
\usepackage[table]{xcolor}
\usepackage{booktabs}
\usepackage{mathtools}
\usepackage{float}
\usepackage{xcolor}
\usepackage[colorlinks=true,allcolors=blue]{hyperref}
\usepackage[capitalise,nameinlink]{cleveref}
\usepackage{tikz}
\usepackage{soul}
\crefname{section}{Sec.}{Figs.}
\graphicspath{{Figures/}}
\begin{document}

\title{Protected hybrid superconducting qubit in an array of gate-tunable Josephson interferometers}
\author{Constantin Schrade}\
\author{Charles M. Marcus}
\address{Center for Quantum Devices, Niels Bohr Institute, University of Copenhagen, 2100 Copenhagen, Denmark}
\author{Andr\'as Gyenis}
\address{Department of Electrical, Computer \& Energy Engineering, University of Colorado Boulder, CO 80309, USA}

\begin{abstract}
We propose a protected qubit based on a modular array of superconducting islands connected by semiconductor Josephson interferometers. The individual interferometers realize effective $\cos2\phi$ elements that exchange `pairs of Cooper pairs' between the superconducting islands when gate-tuned into balance and frustrated by a half flux quantum. If a large capacitor shunts the ends of the array, the circuit forms a protected qubit because its degenerate ground states are robust to offset charge and magnetic field fluctuations for a sizable window around zero offset charge and half flux quantum. This protection window broadens upon increasing the number of interferometers if the individual elements are balanced. We use an effective spin model to describe the system and show that a quantum phase transition point sets the critical flux value at which protection is destroyed.  
\end{abstract}
\maketitle 
A promising route to creating protected qubits is to rely on circuits with underlying symmetries and encode the qubit into distinct eigenstates of the corresponding symmetry operator~\cite{ioffe1998,blatter2001,kalashnikov2020,doucot2012,gyenis2021b, freedman2021}.  Such circuits satisfy the requirements of a protected qubit because (1) the vanishing transition matrix elements between states with different symmetries prevent bit-flip errors, and (2) the near-degeneracy of the qubit states suppresses phase-flip errors. However, a challenge faced by symmetry-protected qubits is the \textit{fragility} of the underlying symmetry. If the relevant symmetry is broken by detuning or noise, protection is lost. Reducing the susceptibility of protected qubit circuits to a symmetry-breaking environment is the focus of this work. 

A paradigmatic symmetry-protected superconducting qubit is the Cooper-pair-parity qubit (or $\cos2\phi$ qubit), which encodes the qubit state onto the parity of the number of Cooper pairs on a superconducting island  \cite{ioffe1998,blatter2001,doucot2012,ioffe2002b,doucot2002,doucot2005,kitaev2006,brooks2013,klots2021}. The ideal $\cos2\phi$ circuit preserves the Cooper-pair parity symmetry via a special type of Josephson junction that only permits the tunneling of pairs of Cooper pairs. Considerable progress has been made towards effectively realizing double-Cooper-pair junctions by a variety of approaches   \cite{bell2014,dempster2014,Groszkowski2018,Paolo2019,gyenis2021a,weiss2019,smith2020,larsen2020}. Improving robustness of the $\cos2\phi$ qubit by constructing arrays that average uncorrelated local noise has also been investigated  \cite{doucot2012,gladchenko2008,ioffe2020,bell2014,plourde2021,dodge2021,shearrow2021}.

\begin{figure}
    \centering
    \includegraphics[width = \columnwidth]{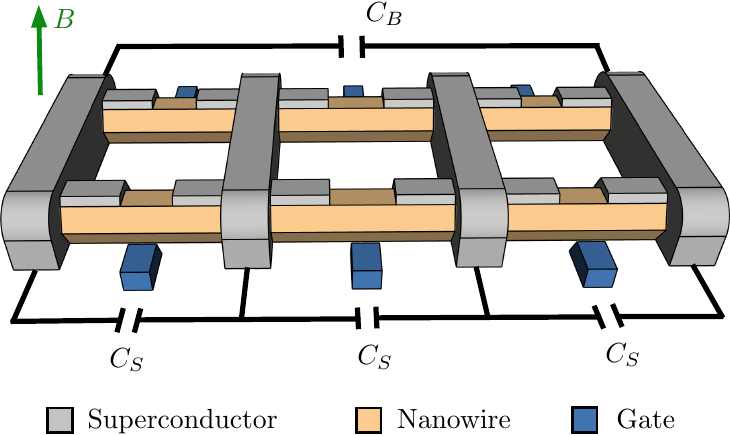}
    \caption{\textbf{Hybrid nanowire-array protected qubit.} The qubit consists of pairs of nanowires (brown) proximitized by superconducting shells and contacted by superconducting islands (gray). The Josephson tunnel barriers are formed by breaking the superconducting shell. In the figure, $N=3$ Josephson interferometers form an array. The two islands at the ends of the array are shunted by a big capacitor with capacitance $C_{B}$, while the islands within the array are coupled by small capacitances $C_{S}\ll C_{B}$. When the Josephson junctions are balanced by local gate electrodes (blue) and the loops are frustrated by magnetic flux around half flux quantum, $\Phi_{0}/2\pm\delta\Phi_\mathrm{ext}$ due to external field $B$, the qubit states are quasi-two-fold degenerate and correspond to charge configurations with opposite total Cooper-pair parity.
    }
    \label{fig1}
\end{figure}

In this paper we introduce and analyze a novel protected qubit based on an array of superconducting islands coupled via semiconductor Josephson interferometers, as illustrated in Fig.\,\ref{fig1}.
By taking advantage of the naturally occurring higher harmonics in semiconductor Josephson junctions \cite{kringhoj2018}, the individual interferometers in the array can readily realize $\cos2\phi$ elements when gate-tuned into balance and frustrated by a half flux quantum \cite{larsen2020}. If a large capacitor shunts the ends of the array, the ground states of the system are two-fold degenerate, carry opposite total Cooper-pair parity, and are robust to offset charge and magnetic field fluctuations. While the array offers protection from noise and offsets that improves with the number of elements, we find that an array as short as two elements considerably improves immunity to flux offset and noise compared to a single interferometer.

Besides protection against energy relaxation as well as fluctuations of charge and flux,  two additional features are noteworthy: (1) The gate-tunability of the semiconductor Josephson junctions allows the interferometers to be tuned into balance using gate voltages rather than additional fluxes. This feature is critical because the proposed concatenation approach enhances the protection only if the elements are pairwise balanced.  (2) The semiconducting-superconducting platform facilitates integration with other hybrid qubits such as gatemons \cite{larsen2015,casparis2018,kringhoj2020} or Majorana qubits \cite{karzig2017,hoffman2016,schrade2018a,schrade2018b}. 

The rest of the paper is organized as follows: In Sec.\,\ref{section1}, we review the $\cos2\phi$ qubit and describe its circuit in terms of a tight-binding model for a particle moving in the $\cos2\phi$ potential. In Sec.\,\ref{section2}, we discuss the realization of a $\cos2\phi$ qubit based on a single semiconductor Josephson interferometer and analyze errors due to flux noise and unbalanced junctions. In Sec.\,\ref{section3}, we introduce the modular array of semiconductor Josephson interferometers and investigate its operation as a protected qubit. We discuss the origin of protection against flux and charge noise, and derive an effective spin Hamiltonian that describes the low-energy spectrum of the array. Finally, in Sec.\,\ref{section4}, we introduce a `giant spin' representation of the effective Hamiltonian, which is useful to compute the critical flux value above which flux noise protection is lost. We associate this critical flux with a quantum phase transition.

\section{The \boldmath$\cos2\phi$ qubit}
\label{section1}
\subsection{Hamiltonian}

\begin{figure}
    \centering
    \includegraphics[width = \columnwidth]{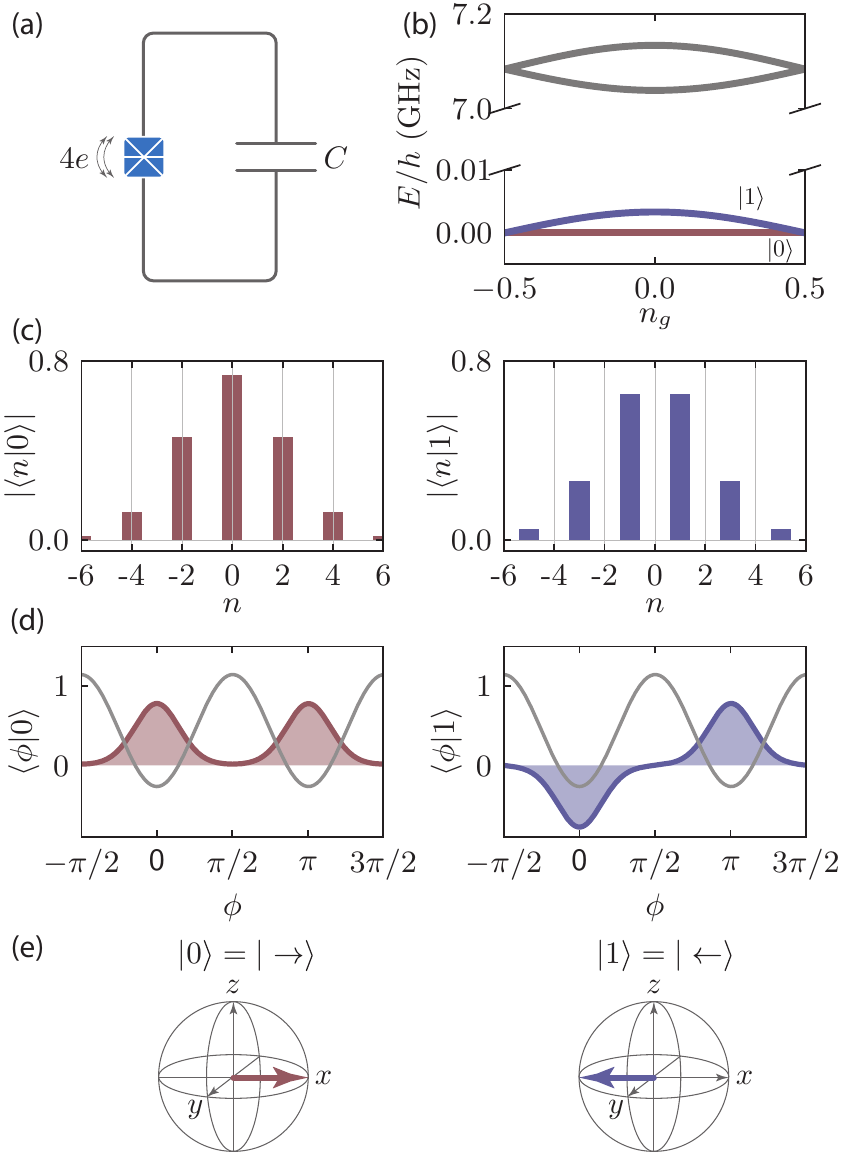}
    \caption{\textbf{Spectrum and eigenstates of the \boldmath$\cos2\phi$ qubit.} (a) Circuit layout of the $\cos2\phi$ qubit with a unique Josephson element that only permits tunneling of `pairs of Cooper-pairs' with charge $4e$ and shunted with capacitance $C$. (b) Energy levels as a function of the gate charge $n_g$. The two lowest energy levels (red and blue) correspond to the two qubit states $\left|0\right\rangle$ and $\left|1\right\rangle$. The higher-lying states (gray) are well-separated from the qubit subspace. (c) Qubit eigenstates in charge space corresponding to superpositions of even charge states for $\left|0\right\rangle$ and odd charge states for $\left|1\right\rangle$. (d) Qubit eigenstates in phase space corresponding to symmetric and antisymmetric combinations of wave functions that are localized in the $0$- and $\pi$-valley of the $\cos2\phi$ Josephson potential (gray). (e) Spin model of the $\cos2\phi$ circuit. The $\left|0\right\rangle$ and $\left|1\right\rangle$ states correspond to spin pointing along the $
    +x$ and $-x$ direction, respectively. Circuit parameters are $E_C/h=0.2$ GHz and $E_{J,2}/h=10$ GHz.
 }
    \label{fig2}
\end{figure}

Before considering the interferometer array, where qubit states are encoded in the overall Cooper-pair parity on multiple islands, we first review the single $\cos2\phi$ qubit. As shown in Fig.\,\ref{fig2}(a), the $\cos2\phi$ qubit consists of a pair of superconducing islands with capacitance $C$ connected via a modified Josephson-like junction that only passes pairs of Cooper pairs (denoted by a Josephson junction symbol with an extra line). The Hamiltonian of this circuit includes a single phase degree of freedom, $\phi$, the superconducting phase difference across the junction,
\begin{equation}
    H_{n_g}=4E_C(n-n_g)^2 - E_{J,2}\cos2\phi,
    \label{eq:H_cos2phi}
\end{equation}
where $n$ is the number operator of Cooper pairs on the capacitor, $E_{J,2}$ is the tunneling amplitude of double Cooper pairs across the junction, $E_C=e^2/2C$ is the charging energy, $e$ is the electron charge, and $n_g$ is the offset charge. We display the wave functions of the ground and first excited states in both charge and phase space in Figs.\,\ref{fig2}(c,d). In charge space, the qubit eigenstates are superpositions of even or odd Cooper-pair parity  states, with an envelope function that broadens with increasing $E_{J,2}/E_C$ ratio, analogous to the transmon qubit \cite{koch2007}. In phase space, qubit states are the symmetric and antisymmetric combinations of states localized in the $0$- or $\pi$-valley of the $\cos2\phi$ potential. The offset charge $n_g$ tunes the qubit transition frequency, which is maximal at $n_g=0$ and zero at $n_g=0.5$, where the qubit states are degenerate, see Fig.\,\ref{fig2}(b).

Protection against energy relaxation in the $\cos2\phi$ qubit results from the symmetry of Cooper-pair parity, which prohibits transitions between the qubit states, $\langle 1|\mathcal{O}|0\rangle=0$ for noise operators $\mathcal{O}$ that do not induce single-Cooper-pair tunneling. The qubit is protected against dephasing because the eigenstates have support over many charge states, $E_{J,2}/E_C\gg1$, leading to exponentially reduced charge dispersion and near-degeneracy. Finally, separation of the qubit states from higher-lying excited states prevents leakage errors.

\subsection{Effective `tight-binding' model}

It is useful to describe the $\cos2\phi$ qubit in terms of a spin model that can shed light on the benefits of using multiple interferometers. We focus on the regime of $E_{J,2}/E_C\gg1$ and use Bloch's theorem to define `atomic' Wannier functions that are localized in the $0$- or $\pi$-valleys of the phase lattice formed by the $\cos2\phi$ potential. 

\begin{figure}
    \centering
    \includegraphics[width = \columnwidth]{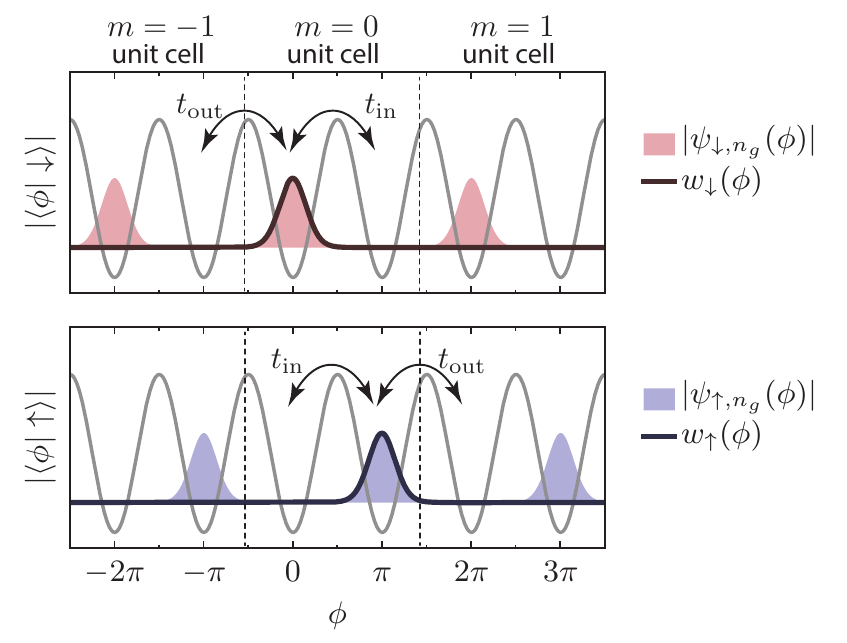}
    \caption{\textbf{Tight-binding representation of the \boldmath$\cos 2\phi$ qubit.} The $\cos 2\phi$ qubit is analogous to a phase particle moving in a periodic lattice formed by the $\cos 2\phi$ Josephson potential (gray). The unit cells consist of two potential minima, the $0$ and $\pi$ valleys. The qubit wave functions are Bloch waves and their symmetric/antisymmetric combinations $\psi_{\uparrow/\downarrow,n_g}(\phi)$ (shown in red and blue) have finite support in one of the two valleys. The Wannier functions $w_{\uparrow/\downarrow}(\phi)$ (shown as black lines) are exponentially localized in a particular unit cell of the lattice. The hopping amplitudes within a unit cell, $t_{\text{in}}$, and outside of the unit cell, $t_{\text{out}}$, are given by the hybridization of neighboring Wannier functions.}
    \label{fig3}
\end{figure}

As a first step, we recall that a qubit with a compact phase degree of freedom is equivalent to a particle moving in a one-dimensional lattice \cite{schon1990, catelani2011,thanh2020}. To show this, we eliminate the offset-charge dependence in $H_{n_g}$ via a unitary rotation, $H=e^{in_g\phi}H_{n_g}e^{-in_g\phi}$, yielding a qubit Hamiltonian of the form,
\begin{equation}
\label{Eq2}
    H=4E_Cn^2 - E_{J,2}\cos2\phi,
\end{equation}
which is equivalent to the Hamiltonian of a particle with mass proportional to $C$ moving in a $V(\phi)=-E_{J,2}\cos2\phi$ potential. The eigenstates of the transformed Hamiltonian $H$ are Bloch waves $\psi_{\ell,n_g}$ that satisfy
quasi-periodic boundary conditions, $\psi_{\ell,n_g}(\phi+2\pi)=e^{in_g2\pi}\psi_{\ell,n_g}(\phi)$ with energy-band-index $\ell$. For the case of the ground and first excited states, we combine the Bloch waves in a symmetric and antisymmetric way, $\psi_{\uparrow/\downarrow,n_g} = (\psi_{0,n_g} \mp \psi_{1,n_g})/\sqrt{2}$. The resulting states are concentrated in either the $0$- or $\pi$-valleys of the extended $\cos2\phi$ potential, see filled curves in Fig.\,\ref{fig3}. Within the spin model, we associate these Bloch states with effective spin states, $\psi_{\uparrow,n_g}(\phi)\Longleftrightarrow \left|\uparrow\right\rangle$ and $\psi_{\downarrow,n_g}(\phi)\Longleftrightarrow \left|\downarrow\right\rangle$. Note that in this spin basis, the qubit eigenstates are symmetric and antisymmetric combinations of spin-up and spin-down states, $\left|0\right\rangle= (\left|\uparrow\right\rangle + \left|\downarrow\right\rangle) / \sqrt{2}$ and $\left|1\right\rangle= (\left|\uparrow\right\rangle - \left|\downarrow\right\rangle) / \sqrt{2}$, see Fig.\,\ref{fig2}(e). 

To derive the Hamiltonian of the spin model, we now introduce Wannier functions on the phase lattice, similar to solid-state systems. These Wannier functions are localized in a particular phase unit cell $m$ and they the are combination of Bloch states at different offset charges,
\begin{equation}
\begin{split}
    w_{s}(\phi-2\pi m)=\frac{1}{\sqrt{M}}\sum_{n_g} e^{-i n_{g}2\pi m}\psi_{s,n_g}(\phi),
\end{split}    
\end{equation}
Here, $s=\uparrow,\downarrow$ labels the effective spin degree of freedom and $M$ is a normalization constant. We depict examples of the Wannier functions in Fig.\,\ref{fig3}.

We next introduce a tight-binding representation of $H$ on the qubit subspace \cite{vanderbilt2018}. First, we project the Hamiltonian $H$ onto the low-energy subspace spanned by $\{\psi_{\uparrow,n_g},\psi_{\downarrow,n_g}\}$. This yields the following matrix elements, 
\begin{equation}
\begin{split}
(H_{\text{eff}})_{ss'}&=\int\mathrm{d}\phi\,\psi_{s,n_g}(\phi)H\psi_{s',n_g}(\phi)\\
&=\sum_{m} e^{i2\pi m n_{g}}\int\mathrm{d}\phi\, 
w_{s}(\phi)Hw_{s'}(\phi-2\pi m).
\label{Eq6}
\end{split}
\end{equation}

Since longer-range hybridizations between Wannier functions in more distant unit cells are negligible ($E_{J,2}/E_{C}\gg1$), we only keep matrix elements that connect Wannier functions that are in the same or nearest-neighbor unit cells. In this tight-binding approximation, there are two types of hybridizations, see Fig.\,\ref{fig3}: tunneling inside a unit cell, $t_{\text{in}}$, and tunneling outside of the unit cell, $t_{\text{out}}$, where
\begin{equation}
\begin{split}
    &t_{\text{in}}=\int \mathrm{d}\phi\, w_\downarrow^*(\phi) H w_\uparrow(\phi),\\
    &t_{\text{out}}=\int \mathrm{d}\phi\, w_\downarrow^*(\phi) H w_\uparrow(\phi+2\pi).
\end{split}
\end{equation}
In case of the $\cos2\phi$ qubit, $t_{\text{in}}=t_{\text{out}}\equiv t<0$. However, we will show below that single Cooper-pair tunneling terms can introduce asymmetries in these tunneling amplitudes.

After collecting all the dominant nearest-neighbor hoppings, we arrive at the following spin Hamiltonian of the $\cos2\phi$ qubit,
\begin{equation}
\begin{split}
\label{Eq6}
   H_{\text{eff}} & \approx2t\cos(\pi n_g)\left\{\cos(\pi n_g)\sigma_x + \sin(\pi n_g)\sigma_y\right\}= \\
   & =2t\cos(\pi n_g)\tilde\sigma_{x},
  \end{split}
\end{equation}
where we have introduced the offset-charge dependent rotated Pauli matrices $\tilde\sigma_{x}=R_{z}^{\dag}\sigma_{x}R_{z}$ with $R_{z}=e^{-i\pi n_{g}\sigma_{z}/2}$. At $n_g=0$, we denote the ground state of the spin model in Eq.\,\eqref{Eq6}, as $\left|0\right\rangle=\left|\rightarrow\right\rangle$ and the first-excited state as $\left|1\right\rangle= \left|\leftarrow\right\rangle$, indicating that these states point along the  $+x$ and $-x$ spin direction, see Fig.\,\ref{fig2}(e). In terms of the Cooper-pair parity, the $\left|\rightarrow\right\rangle$ corresponds to the even, while the $\left|\leftarrow\right\rangle$ corresponds to the odd parity state. Away from $n_g=0$, the two eigenstates still point in opposite direction in the $x$$y$ plane but the tunnel splitting is reduced. The transition energy of the qubit is $E_{01}=\left|4t\cos(\pi n_g)\right|$, which is in agreement with the exact result shown in Fig.\,\ref{fig2}(b).

\section{Single-interferometer qubit}
\label{section2}

\subsection{Josephson energy}

We now turn to the realization of a single $\cos2\phi$ qubit using nanowire Josephson junctions, as discussed recently in Ref.~\cite{larsen2020}. In contrast to conventional Al/AlO$_x$/Al Josephson junctions with Josephson energy $E_{J}(\phi)= -E_{J,1}\cos\phi$, the Josephson energy of semiconductor nanowire junctions contains higher harmonics, $E_{J}(\phi)= -E_{J,1}\cos\phi + E_{J,2}\cos2\phi+\dots$. These higher harmonics originate from a few high-transmission channels that mediate the Cooper-pair tunneling via Andreev bound states in the junction. For junctions shorter than the superconducting coherence length, this yields a Josephson energy \cite{beenakker1991,martinis2004,kringhoj2018}, 
\begin{equation}
E_{J}(\phi) = -\Delta\sum_m\sqrt{1-T_m\sin^2{\phi/2}}.
\label{Eq7}
\end{equation}
Here, the transmission coefficient $T_m$  characterizes the $m$-th Andreev bound state and $\Delta$ is the superconducting gap. Although the second harmonic, $E_{J,2}\cos2\phi$, can be sizable in a few-channel junctions \cite{larsen2020}, the first harmonic, $-E_{J,1}\cos\phi$, remains the dominant contribution the Josephson energy unless deliberately removed.

\begin{figure}
    \centering
    \includegraphics[width = \columnwidth]{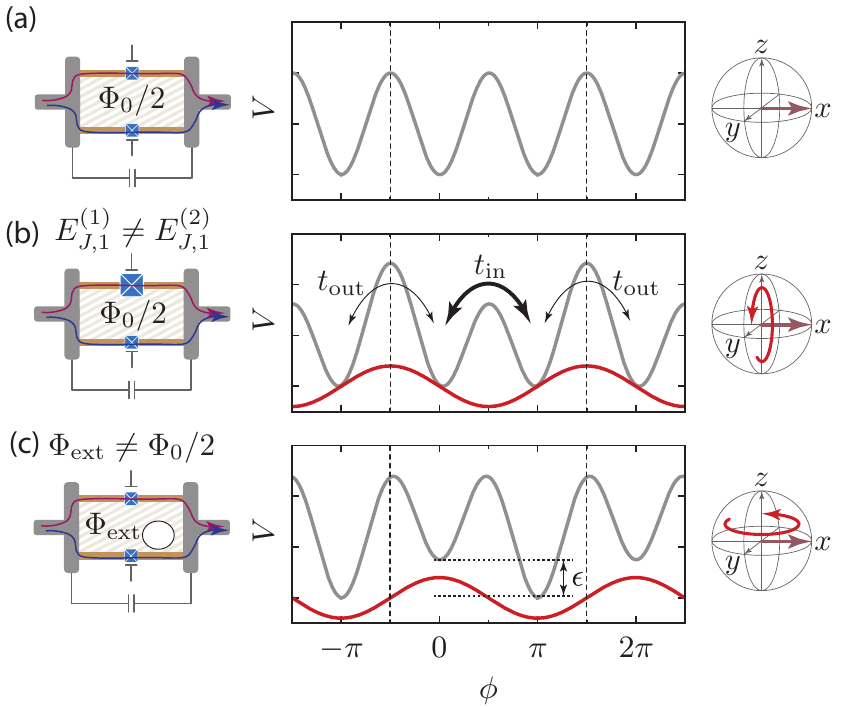}
    \caption{\textbf{Single-interferometer qubit and possible error sources} (a) Left panel: Single-loop qubit frustrated by a half-flux quantum (dashed area) with two highly-transmissive Josephson junctions that are gate-tuned into balance. Red and blue long arrows show the two possible tunneling paths for Cooper pairs. Middle panel: $\cos2\phi$ Josephson potential of the single-loop qubit. Right panel: spin model of the qubit, where the eigenstates correspond to spins pointing along the $+x$/$-x$ direction. (b) Left panel: Single-loop qubit with unbalanced junctions. Middle panel: Resulting Josephson potential with a correction $\propto\sin\phi$ that induces different barrier heights and different inter- and intra-unit cell tunneling amplitudes, $t_{\text{in}}\neq t_{\text{out}}$. Right panel: unbalanced junctions induce a qubit rotation around the $x$-axis. (c) Left panel: Single-loop qubit with a flux bias away from the optimal point, $\Phi_\mathrm{ext}\neq\Phi_{0}/2$.  Middle panel: The corresponding Josephson potential with a correction $\propto\cos\phi$ that induces a different on-site potential for the $0$/$\pi$ valleys. 
Right panel: Flux bias error causes a qubit rotation around the $z$-axis. }
    \label{fig4}
\end{figure}

One approach to ensuring that the leading term in the Josephson energy is the second harmonic is to make use of Aharonov-Bohm interference \cite{bell2014,larsen2020} by considering a \textit{symmetric} Josephson interferometer comprised of two semiconductor Josephson junctions enclosing a flux $\Phi_\mathrm{ext}$, see Fig.\,\ref{fig4}(a). When Cooper pairs follow the two different paths of the interferometer, they acquire a relative Aharonov-Bohm phase, $2\pi q\Phi_\mathrm{ext}/h$, proportional to the charge $q$. At one half quantum of external flux, $\Phi_\mathrm{ext}=\Phi_{0}/2=h/4e$, single Cooper-pairs interfere destructively with a relative Aharonov-Bohm phase $\pi$ because they carry charge $q=2e$. In contrast, double Cooper pairs interfere constructively with relative Aharonov-Bohm phase $2\pi$ because they carry charge $q=4e$. As a result, the leading term in the Josephson energy is given by double Cooper pair hopping and the circuit realizes an effective $\cos2\phi$ element.

To formalize this argument, we note that the Josephson energies of the individual junctions are well-approximated by $E_{J}^{(k)}(\phi)\approx -E_{J,1}^{(k)}\cos\phi + E_{J,2}^{(k)}\cos2\phi$ with $k=1,2$ labelling the interferometer arms. In the presence of an external flux $\Phi_\mathrm{ext}$, the total Josephson energy of the circuit reads, 
\begin{equation}
    E_{J}^\mathrm{(tot)}(\phi) = E_{J}^{(1)}(\phi-\pi\Phi_\mathrm{ext}/\Phi_\mathrm{0}) + E_{J}^{(2)}(\phi+\pi\Phi_\mathrm{ext}/\Phi_\mathrm{0}).
\end{equation}
After substitution, we see that when the first-order terms are equal, $E^{(1)}_{J,1}=E^{(2)}_{J,1}$, and the flux is half flux quantum, $\Phi_\mathrm{ext}=\Phi_0/2$, single Cooper-pair tunneling vanishes, and $E_{J}^\mathrm{(tot)}(\phi)=(E_{J,2}^{(1)}+E_{J,2}^{(2)})\cos2\phi$.

\subsection{Error sources}

To achieve complete suppression of single Cooper-pair tunneling events (complete destructive interference), the circuit needs to satisfy \textit{two} requirements. First, the external flux through the loop needs to be exactly biased at half flux quantum, $\Phi_\mathrm{ext}=\Phi_{0}/2$. Second, the single Cooper-pair tunneling amplitudes across the two junctions needs to be equal, $E^{(1)}_{J,1}=E^{(2)}_{J,1}$. If the circuit fails to satisfy any of these two requirements, there will be a finite tunneling contributions of single Cooper pairs across the device. Although in both cases the Cooper-pair parity protection is destroyed, the two types of errors are fundamentally different, and protection against them requires different strategies. 

The first type of errors results from unbalanced transmission amplitudes, $\delta E_{J,1}=E_{J,1}^{(1)}-E_{J,1}^{(2)}\neq0$, see Fig.\,\ref{fig4}(b). This imbalance leads to a sinusoidal error contribution in the Josephson energy, $\delta E_{J}(\phi) = \delta E_{J,1} \sin\phi$. This type of error changes the tunnel barrier between $0$- and the $\pi$-valleys without inducing a tilt between the two valleys. As a result, the inter- and intra-unit cell hybridizations of the Wannier functions become asymmetric. In the modified spin-Hamiltonian, this asymmetric hybridization leads to $\sigma_{x}$-type errors, 
\begin{equation}
   \delta H_{\text{eff}}=(t_\text{in} - t_\text{out})\sigma_x.
\end{equation}

The second type of errors result from noise or offset in magnetic flux away from one half flux quantum, see Fig.\,\ref{fig4}(c). For balanced junctions, $E^{(1)}_{J,1}=E^{(2)}_{J,1}\equiv E_{J,1}$, but flux detuned from half flux quantum, $\Phi_\mathrm{ext}=\Phi_0/2+\delta\Phi_\mathrm{ext}$ with $\delta\Phi_\mathrm{ext}/\Phi_0\ll 1$, the Josephson energy acquires a cosinusoidal error contribution, $\delta E_{J}(\phi) = -2E_{J,1}(\delta\Phi_\mathrm{ext}/\Phi_0) \cos\phi$. Such an error induces a tilt between the $0$- and the $\pi$-valleys, i.e., different on-site energies for the Wannier functions $w_{\uparrow}(\phi)$ and $w_{\downarrow}(\phi)$. In the spin model, these different on-site energies lead to a $\sigma_z$ error term, 
\begin{equation}
\label{Eq9}
   \delta H_{\text{eff}}=-\epsilon\sigma_z.
\end{equation}
Here, the amplitude of the $\sigma_z$ term is determined by the flux detuning away from half flux quantum, $\epsilon\approx E_{J,1}(\delta\Phi_\mathrm{ext}/\Phi_0)$.

In the following, we will show that the flux errors can be effectively eliminated with multi-interferometer $\cos2\phi$ qubits, while errors due to unequal junction transmissions can be prevented by in-situ gate-tuning of the junctions.

\begin{figure*}
    \centering
    \includegraphics[width = \textwidth]{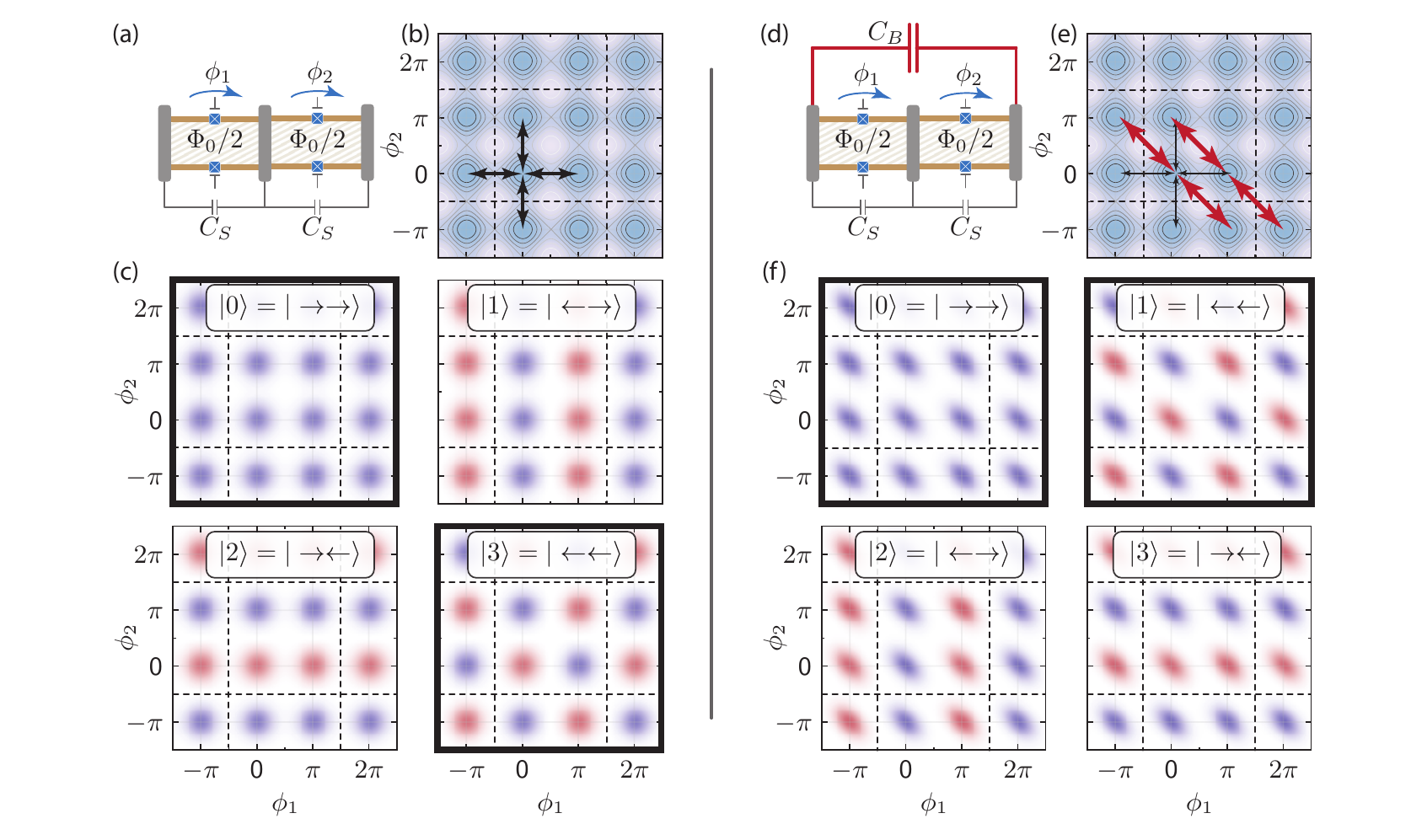}
    \caption{\textbf{Two-interferometer qubit and its eigenstates} (a) The two-interferometer qubit with only neighboring islands coupled with small capacitances $C_{S}$. (b) Two-dimensional Josephson potential as a function of the superconducting phase differences $\phi_{1}$ and $\phi_{2}$ across the junctions. The dominant hybridizations are between Wannier states that are located in nearest-neighboring valleys along the $\phi_{1}$ and $\phi_{2}$ axes (black arrows). (c) Bloch wave functions of the four lowest-lying states. The ferromagnetic configurations (highlighted with wide boundaries), the ground state, $|0\rangle=\left|\rightarrow \rightarrow\right\rangle$ and the third excited state, $|3\rangle=\left|\leftarrow \leftarrow\right\rangle$, do not form a separate qubit subspace. (d) In the protected regime, the superconducting islands at the ends of the array are coupled by a big shunt capacitor with capacitance $C_{B}\gg C_{S}$. (e) The hybridization between Wannier states located along the diagonal direction is enhanced in this protected configuration (red arrows). (f) The Bloch wave functions of the four lowest-lying states. The ferromagnetic configurations, $|0\rangle=\left|\rightarrow \rightarrow\right\rangle$ and $|1\rangle=\left|\leftarrow \leftarrow\right\rangle$, now form a nearly degenerate subspace and can be used for encoding a protected qubit. }
    \label{fig5}
\end{figure*}

\section{Multi-interferometer qubits}
\label{section3}

\subsection{Hamiltonian}

With the motivation of reducing the effect of flux offset and noise ($\sigma_z$ errors),  we now extend our spin model to multi-interferometer qubits (see Fig.\,\ref{fig1} for the case of three connected interferometers). Each pair of junctions in the interferometers is gate-tuned into balance and the interferometer loops are frustrated with half-flux quanta. The superconducting islands of the array are capacitively coupled, such that the islands within the array are coupled by a small capacitance $C_{S}$, while the two islands at the ends of the array are coupled by a big capacitance $C_{B}\gg C_{S}$. The Hamiltonian of the setup with $N$ interference loops is, 
\begin{equation}
\begin{split}
H^{(N)}_{\boldsymbol{n}_{g}}=
&\sum_{i,j=1}^{N}
4E^{(ij)}_{C}
(n_{i}-n^{(i)}_{g})
(n_{j}-n^{(j)}_{g})
\\
-&
\sum_{i=1}^{N} 
E^{(i)}_{J,2}
\cos2\phi_{i}.
\end{split}
\label{Eq11}
\end{equation}
Here, $n_{i}$ is the charge operator with offset charge $n^{(i)}_{g}$ of the $i$-th loop, $\phi_{i}$ is the conjugate phase operator, $E^{(i)}_{J,2}$ is the double Cooper-pair tunneling amplitude, and the charging energies are $E^{(ij)}_{C}=e^{2}C^{-1}_{ij}/2$, where $C_{ij}$ is the capacitance matrix (see Appendix A for details).

Similar to the single-interferometer case, we can eliminate the offset charge dependence of the Hamiltonian in Eq.\,\eqref{Eq11} by a unitary rotation, $H^{(N)}=e^{i\boldsymbol{n}_{g}\cdot\boldsymbol{\phi_{i}}}H^{(N)}_{\boldsymbol{n}_{g}}e^{-i\boldsymbol{n}_{g}\cdot\boldsymbol{\phi_{i}}}$ with $\boldsymbol{n}_{g}=(n^{(1)}_{g},\dots,n^{(N)}_{g})^{T}$ and $\boldsymbol{\phi}=(\phi_{1},\dots,\phi_{N})^{T}$, yielding
\begin{equation}
\begin{split}
H^{(N)}=
&\sum_{i,j=1}^{N}
4E^{(ij)}_{C}
n_{i}
n_{j}
-
\sum_{i=1}^{N} 
E^{(i)}_{J,2}
\cos2\phi_{i}.
\end{split}
\label{Eq12}
\end{equation}
The offset charge dependence appears in the boundary conditions on the eigenfunctions $\psi_{\ell,\boldsymbol{n}_{g}}(\boldsymbol{\phi})$, which are Bloch waves with quasi-periodicity, $\psi_{\ell,\boldsymbol{n}_g}(\boldsymbol{\phi}+2\pi\boldsymbol{e}_{i})=e^{i2\pi\boldsymbol{e}_{i}\cdot\boldsymbol{n}_{g}}\psi_{\ell,\boldsymbol{n}_g}(\boldsymbol{\phi})$ where $\boldsymbol{e}_{i}=(0,\dots,1_{i},\dots,0)^{T}$. 

\subsection{Two-interferometer qubit}

We can gain insight into the origin of protection against flux errors by examining the simplest extension, with $N=2$ interferometer loops. In this case, the Josephson energy of the setup, $-E^{(1)}_{J,2}\cos2\phi_1-E^{(2)}_{J,2}\cos2\phi_2$, has four minima in the unit cell at positions $\{(0,0),(0,\pi),(\pi,0),(\pi,\pi)\}$. We denote the linear combinations of the four lowest-energy Bloch waves with support in each of the four minima by $\{\psi_{\uparrow\uparrow,\boldsymbol{n}_{g}},\psi_{\uparrow\downarrow,\boldsymbol{n}_{g}},\psi_{\downarrow\uparrow,\boldsymbol{n}_{g}},\psi_{\downarrow\downarrow,\boldsymbol{n}_{g}}\}$, and use them to define Wannier functions that are localized in a particular valley of the two-dimensional phase lattice,
\begin{align}
\hspace{-4pt}
    w_{s_{1}s_{2}}(\boldsymbol{\phi}-2\pi \boldsymbol{m})=\frac{1}{\sqrt{M}}\sum_{\boldsymbol{n}_g} e^{-i 2\pi \boldsymbol{m}\cdot\boldsymbol{n}_{g}}\psi_{s_{1}s_{2},\boldsymbol{n}_g}(\boldsymbol{\phi}).
\label{Eq13}
\end{align}
Here, $s_{1},s_{2}=\uparrow,\downarrow$ denote the effective spin degrees of freedom, $M$ is a normalization factor, and $\boldsymbol{m}=(m_{1},m_{2})$ is a two-dimensional matrix of integers. 
 
After projecting the Hamiltonian of Eq.\,\eqref{Eq11} onto the $\{\psi_{s_1s_2,\boldsymbol{n}_{g}}\}$-subspace and expressing the matrix elements in terms of these Wannier functions, we obtain an effective tight-binding Hamiltonian,
\begin{equation}
\label{Eq14}
(H^{(2)}_{\text{eff}})_{s_1s_2,s_3s_4}
=
\sum_{\boldsymbol{m}} e^{i2\pi \boldsymbol{m}\cdot\boldsymbol{n}_{g}}\,
t_{s_1s_2,s_3s_4}(\boldsymbol{m}),
\end{equation}
where
\begin{equation}
t_{s_1s_2,s_3s_4}(\boldsymbol{m})
=\int\mathrm{d}\boldsymbol{\phi}\, 
w_{s_{1}s_{2}}(\boldsymbol{\phi})H^{(2)}w_{s_{3}s_{4}}(\boldsymbol{\phi}-2\pi \boldsymbol{m}).\nonumber
\end{equation}
In this representation, we can associate the Bloch wave functions $\{\psi_{s_1s_2,\boldsymbol{n}_{g}}\}$ with the states of a double spin-1/2 system $\{\left|s_1s_2\right\rangle\}$ and the interaction between the spins with the hybridizations $t_{s_1s_2,s_3s_4}(\boldsymbol{m})$ between Wannier functions. We consider two limiting cases, $C_{B}\ll C_{S}$ and $C_{B}\gg C_{S}$ and focus on the situation $n^{(i)}_{g}=0$ for simplicity. We will discuss the effects of offset charges later.

For $C_{B}\ll C_{S}$ coupling between the ends of the array is negligible, $E^{(12)}_{C}\approx0$, see Fig.\,\ref{fig5}(a). In the Wannier picture, this implies that nearest-neighbor hopping along the $\phi_{1}$- and $\phi_{2}$-direction 
give the dominant contributions to the tight-binding Hamiltonian, see Fig.\,\ref{fig5}(b). By retaining only such nearest-neighbor hybridizations, the tight-binding Hamiltonian for $C_{B}\ll C_{S}$ takes on the simplified form, 
\begin{equation}
   H^{(2)}_{\text{eff}}\approx 2t\,(\sigma^{(1)}_x+\sigma^{(2)}_x),
\label{Eq15}
\end{equation}
which describes two separate spin-1/2 systems with no spin-spin interaction. The eigenstates and energy levels are [Fig.\,\ref{fig5}(c)],
\begin{equation}
\label{Eq16}
\begin{split}
&\left|\rightarrow \rightarrow\right\rangle
\propto
(\left|\uparrow\right\rangle^{(1)}+\left|\downarrow\right\rangle^{(1)})
(\left|\uparrow\right\rangle^{(2)}+\left|\downarrow\right\rangle^{(2)}), 
\quad
E_{\rightarrow \rightarrow}=0
\\
&\left|\leftarrow\rightarrow\right\rangle
\propto
(\left|\uparrow\right\rangle^{(1)}-\left|\downarrow\right\rangle^{(1)})
(\left|\uparrow\right\rangle^{(2)}+\left|\downarrow\right\rangle^{(2)}), 
\quad
E_{\leftarrow \rightarrow}=4t
\\
&\left|\rightarrow\leftarrow\right\rangle
\propto
(\left|\uparrow\right\rangle^{(1)}+\left|\downarrow\right\rangle^{(1)})
(\left|\uparrow\right\rangle^{(2)}-\left|\downarrow\right\rangle^{(2)}), 
\quad
E_{\rightarrow\leftarrow}=4t
\\
&\left|\leftarrow \leftarrow\right\rangle
\propto
(\left|\uparrow\right\rangle^{(1)}-\left|\downarrow\right\rangle^{(1)})
(\left|\uparrow\right\rangle^{(2)}-\left|\downarrow\right\rangle^{(2)}), 
\quad
E_{\rightarrow \rightarrow}=8t
\end{split}
\end{equation}
Here, the ground state, $\left|\rightarrow \rightarrow\right\rangle$, and the highest energy state, $\left|\leftarrow \leftarrow\right\rangle$, are both ferromagnetic with both spins pointing along the $+x$ or $-x$ direction, and correspond to states of opposite total Cooper-pair parity. These two configurations are good candidates for encoding a protected qubit since local magnetic field errors ($\sigma^{(i)}_{z}$ error terms) are unable to mix them, $\left\langle\rightarrow \rightarrow\right|\sigma^{(i)}_{z}\left|\leftarrow \leftarrow\right\rangle=0$. Thus, the suggested qubit encoding scheme provides protection against flux noise. However, in the limit of $C_{B}\ll C_{S}$, the ferromagnetic configurations do not form a low-energy subspace that is well-separated from the remaining states. Hence, the qubit encoding in the $\{\left|\rightarrow \rightarrow\right\rangle,\left|\leftarrow \leftarrow\right\rangle\}$-subspace requires some modification. 

\begin{figure}
    \centering
    \includegraphics[width = \columnwidth]{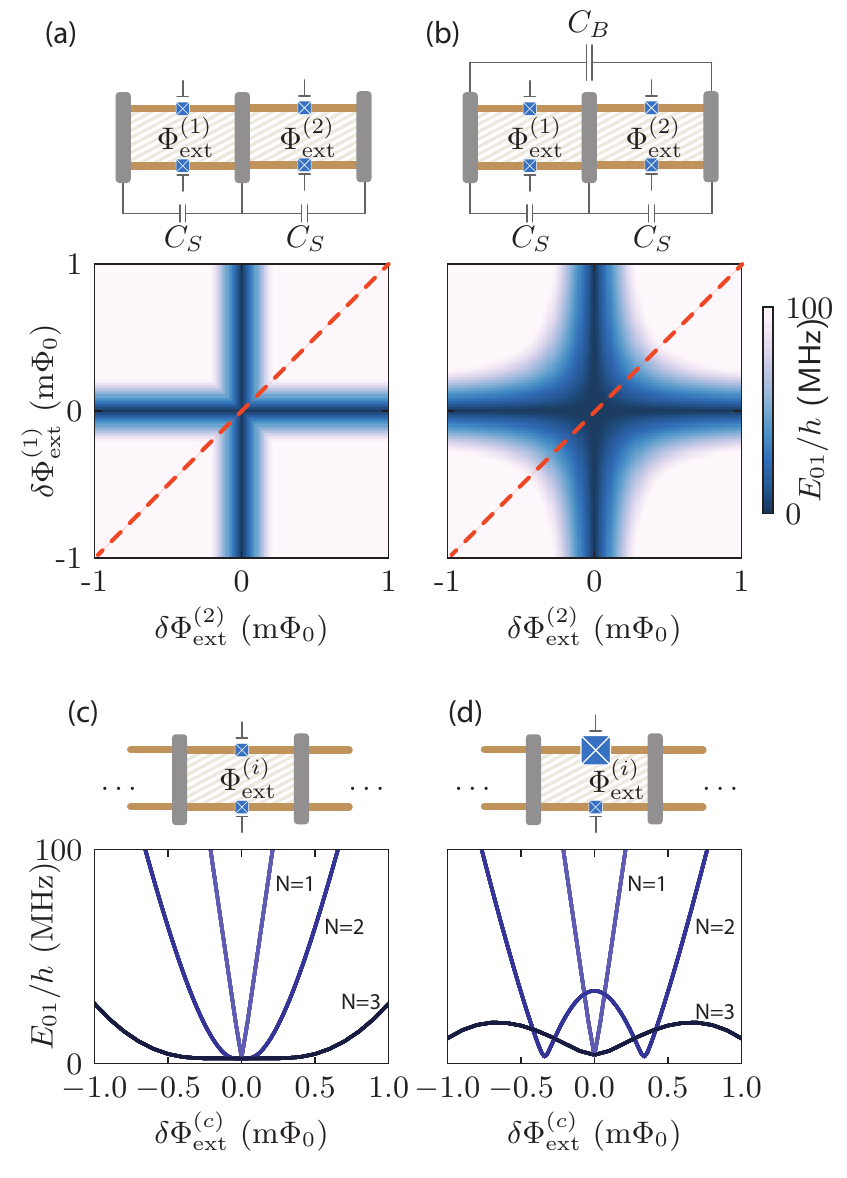}
    \caption{\textbf{Flux-noise protection in arrays} (a) Example for the energy dispersion $E_{01}(\delta\Phi_\mathrm{ext}^{(1)},\delta\Phi_\mathrm{ext}^{(2)})/h$ for an unprotected two-interferometer qubit with $C_{B}=0$, where $\delta\Phi_\mathrm{ext}^{(i)}$  are the flux offsets from half flux quantum in the two loops. The ground state degeneracy exists only if $\delta\Phi_\mathrm{ext}^{(1)}=0$ or $\delta\Phi_\mathrm{ext}^{(2)}=0$. (b) For $C_{B}\gg C_{S}$, the qubit remains protected within a broader window around half flux quantum. (c) Energy dispersion $E_{01}/h$ for an array with $N=1,2,3$ symmetric interferometer loops as a function of the correlated (global) deviation $\delta\Phi_\mathrm{ext}^{(c)}\equiv\delta\Phi_\mathrm{ext}^{(i)}$ away from half flux quantum. This is the direction in flux space, where flux noise has the most detrimental effect on the qubit degeneracy [dashed lines on (a)-(b)]. The protection window broadens upon appending additional loops to the interferometer array. (d) In case of unbalanced junctions, the protection window against flux noise is absent. The capacitance values are $C_B=100$ fF ($N=1$), $C_S=10$ fF, $C_B=200$ fF ($N=2$), $C_S=10$ fF, $C_B=350$ fF ($N=3$), the tunneling energy is $E_{J,2}/h=10$ GHz; the slope of the asymmetry term due to flux noise is $E_{J,1}/\Phi_0=$ 250 GHz/$\Phi_0$ (similar to \cite{larsen2020}); the asymmetry term due to unbalanced junctions in (d) is $E_{J,1}/h=2$ GHz.}
    \label{fig6}
\end{figure}

One modification that leads to the desired arrangement, namely the two ferromagnetic configurations becoming a low lying pair of nearly degenerate states, can be realized by adding a large capacitive coupling between the end two islands, $C_{B}\gg C_{S}$, see Fig.\,\ref{fig5}(d). In the Wannier picture, this enhances  diagonal next-nearest-neighbor hybridization [Fig.\,\ref{fig5}(e)], 
\begin{equation}
   H^{(2)}_{\text{eff}}\approx 2t\,(\sigma^{(1)}_x+\sigma^{(2)}_x) -2J\,\sigma^{(1)}_x\sigma^{(2)}_x .
\label{Eq17}
\end{equation}
Here, the coupling $J\equiv -t_{\uparrow\downarrow,\downarrow\uparrow}(0)>0$ arises from the elongated Wannier function overlap along one of the diagonals, see derivation in Appendix B. For $J\gg |t|$, the two ferromagnetic states $\{\left|\rightarrow \rightarrow\right\rangle,\left|\leftarrow \leftarrow\right\rangle\}$ form a pair of well-separated low-energy states protected against flux noise. In terms of qubit energy flux dispersion, this manifests as a broadening of the `sweet spot' around $\Phi_\mathrm{ext}=\Phi_0/2$ as shown in Fig.\,\ref{fig6}.

Including offset charges (Appendix B) generalizes the coupling term in Eq.\,\eqref{Eq17} in the $J\gg |t|$ regime to
\begin{equation}
-2J\,\sigma^{(1)}_x\sigma^{(2)}_x\longrightarrow-2J\cos(\pi[n^{(1)}_g-n^{(2)}_g])\,\tilde\sigma^{(1)}_x\tilde\sigma^{(2)}_x, 
\label{Eq18}
\end{equation}
where $\tilde\sigma^{(i)}_{x}$ are the rotated Pauli matrices defined above. Thus, the finite offset charges have two effects. First, the offset charges induce a spin-rotation around the $z$-axis by an angle $\pi n^{(i)}_{g}/2$. Critically, in the rotated spin basis, the ferromagnetic configurations remain protected from mixing due to local $\sigma^{(i)}_{z}$ error because $\tilde\sigma^{(i)}_{z}=(R^{(i)}_{z})^{\dag}\sigma^{(i)}_{z}R^{(i)}_{z}=\sigma^{(i)}_{z}$. Second, the offset charges give rise to a modulation of the coupling $J$ by a factor of $\cos(\pi[n^{(1)}_g-n^{(2)}_g])$. Thus, the ferromagnetic configuration remains the ground states as long as $|n^{(1)}_g-n^{(2)}_g|$ is small, see Appendix C. 

A requirement for realizing the ferromagnetic eigenstates, and thus the robust parity protected states, is the balanced transmission amplitudes in both loops. If the single Cooper-pair tunneling amplitudes are unbalanced, additional $\sigma_y^{(1)}\sigma_y^{(2)}$ and $\sigma_z^{(1)}\sigma_z^{(2)}$  coupling terms arise. Such coupling terms introduce mixing between the ferromagnetic states and destroy the protection, as, for example, $\left\langle\rightarrow \rightarrow\right|\sigma^{(1)}_{z}\sigma^{(2)}_{z}\left|\leftarrow \leftarrow\right\rangle\neq0$. This leads to that the enlarged sweet spot disappears; see Fig.\,\ref{fig6}(d), which shows the flux dependence of the eigenstates in the presence of unbalanced junctions. This requirement further highlights the benefits of semiconductor-based tunnel junctions, where the junction transmission amplitudes are \textit{in-situ} tunable.

\subsection{Multi-interferometer qubit}
Finally, we now comment on the extension to $N\geq2$ unit cells in the array.  For this generalized scenario, we again focus on the case of zero offset charges, and interpret Eq.\,\eqref{Eq12} as the Hamiltonian of a particle hopping between the $2^{N}$ minima of the potential $-\sum_{i=1}^{N}E^{(i)}_{J,2}\cos2\phi_{i}$ with an inverse effective mass tensor $E^{(ij)}_{C}$. The inverse effective mass of the particle along the $\phi_{i}$ axes and the diagonals are uniform, $E^{(ij)}_{C}\equiv E^{\text{(diag)}}_{C}>0$ for $i=j$ and $E^{(ij)}_{C}\equiv E^{\text{(off-diag)}}_{C}<0$ for $i\neq j$ (see Appendix A). Hence, if we retain only nearest neighbor hopping along the main axes and next-nearest neighbor hopping along the diagonals, the effective tight-binding Hamiltonian takes the form,
\begin{equation}
\label{Eq19}
H^{(N)}_{\text{eff}}\approx 2t\sum_{i=1}^{N}\sigma^{(i)}_{x}
-
\frac{2J}{N}\sum_{i<j}^{N}
\sigma^{(i)}_{x}
\sigma^{(j)}_{x}
-
\sum^{N}_{i=1}\epsilon_{i}\sigma^{(i)}_{z}.
\end{equation}
Here, $t$ and $J$ denote the overlap integrals for nearest and next-nearest neighbor Wannier functions, while $\epsilon_{i}$ are the flux offsets. The factor $1/N$ in the term $\propto J$ ensures that the ground state energy per effective spin remains finite in the thermodynamic limit.

Such array exhibits a near-ground state degeneracy for a sizable window around half flux quantum that broadens upon increasing the length $N$ of the array, see Fig.\,\ref{fig6}. This \textit{robust} degeneracy for the multi-interferometer devices with $N>1$ is notably different from the \textit{fragile} degeneracy for the single-interferometer device with $N=1$ and permits the encoding of a protected qubit. In terms of the effective model in Eq.\,\eqref{Eq19}, the degeneracy implies that the ferromagnetic coupling $J$ remains the dominant energy scale also for longer arrays provided that $C_{B}\gg C_{S}$. In the next section, we will derive a specific upper bound for the flux noise protection window in terms of the effective parameters $(t,J,\epsilon_{i})$.

\section{`Giant spin' representation}
\label{section4}
We next use the above results to derive a specific upper bound for the protection window against flux noise and offset. We find that the critical flux at which protection is destroyed corresponds to a quantum phase transition. 

\subsection{Hamiltonian}

To begin, note that the Hamiltonian in Eq.\,\eqref{Eq19} involves an all-to-all interaction between effective spins proportional to $\sigma^{(i)}_{x}\sigma^{(j)}_{x}$. This allows us to introduce an $N$-spin representation (or `giant spin') $S_{x,y,z}=\sum_{i=1}^{N}\sigma^{(i)}_{x,y,z}/2$ with $[S_{\alpha},S_{\beta}]=i\varepsilon_{\alpha\beta\gamma}S_{\gamma}$ and rewrite the Hamiltonian in the following form,
\begin{equation}
\label{Eq20}
H^{(N)}_{\text{eff}}
=
-2(
\epsilon S_{z}
-
2t\,
S_{x}
)
-
(4J/N)
S_{x}^{2}
.
\end{equation}
Here, terms proportional to $S_{x}$ and $S_{z}$ describe effective magnetic fields along the $x$- and $z$-direction, while the term proportional to $S^{2}_{x}$ describes a magnetic `easy-axis' along the $x$-direction. In comparison to Eq.\,\eqref{Eq19}, we consider correlated flux noise or offset, $\epsilon\equiv\epsilon_{i}$, due to fluctuations of the global magnetic field. 
The Hamiltonian written in the form of Eq.\,\eqref{Eq20} with $t=0$ is the Lipkin-Meshkov-Glick Hamiltonian \cite{lipkin1,lipkin2,lipkin3}. In the following, we analyze this Hamiltonian through the lens of the interferometer-array protected qubit.

\subsection{Symmetries}

We begin with a discussion on the symmetries of the Hamiltonian in Eq.\,\eqref{Eq20} and the structure of its eigenstates. 

First, the Hamiltonian preserves the magnitude
 of the total giant spin $\boldsymbol{S}=(S_{x},S_{y},S_{z})$ so that,
 \begin{equation}
[
H^{(N)}_{\text{eff}}, \boldsymbol{S}^{2}
]=0,
\end{equation}
This symmetry implies that the eigenstate sectors with different total giant spin $S$ decouple. Consequently, we can focus on the $S=N/2$ sector that contains the low-energy states of our system. 

Second, the Hamiltonian with $t=0$ has also a spin-flip symmetry, prohibiting the coupling between states with a different number of spins pointing along the $z$-direction,
\begin{equation}
\left[
H^{(N)}_{\text{eff}}, \prod_{i=1}^{N}\sigma^{(i)}_{z}
\right]=0
\quad
,
\quad
(t=0)
\end{equation}
In a common eigenbasis of $H^{(N)}_{\text{eff}}$ and $\prod_{i=1}^{N}\sigma^{(i)}_{z}$, this symmetry implies
$\langle S_{x}\rangle=\langle S_{y}\rangle=0$ and $\langle S_{x}S_{z}\rangle=\langle S_{y}S_{z}\rangle=0$.

\subsection{Phase diagram}

We next derive a phase diagram for the Hamiltonian of Eq.\,\eqref{Eq20} that characterizes the parameter space regions for which the ground state subspace is two-fold degenerate and permits the encoding a protected qubit. We initially focus on the limit when the two ends of the device are shunted by a large capacitor, $C_{B}\gg C_{S}$. This ensures that the effective magnetic field $\propto t$ in Eq.\,\eqref{Eq20} is small compared to the ferromagnetic coupling $\propto J$ and the flux noise $\propto \epsilon$. Moreover, we will also assume the limiting case of a device with many interference loops, $N\gg 1$. This large-$N$ limit allows us to approximate the ground states by `spin coherent states' of the form \cite{dusuel2005}, 
\begin{equation}
\label{Eq23}
|\theta,\chi\rangle=\bigotimes^{N}_{i=1}
\left[
\cos\left(\frac{\theta}{2}\right)
\left|\uparrow\right\rangle^{(i)}
+
e^{i\chi}
\sin\left(\frac{\theta}{2}\right)
\left|\downarrow\right\rangle^{(i)}
\right],
\end{equation}
with mean spin direction that is given by a point on the unit sphere, $\boldsymbol{n}_{0}=\langle\theta,\chi|\boldsymbol{S}|\theta,\chi\rangle=(\sin\theta\cos\chi,\sin\theta\sin\chi,\cos\theta)$.

\begin{figure}
    \centering
    \includegraphics[width = 0.9\columnwidth]{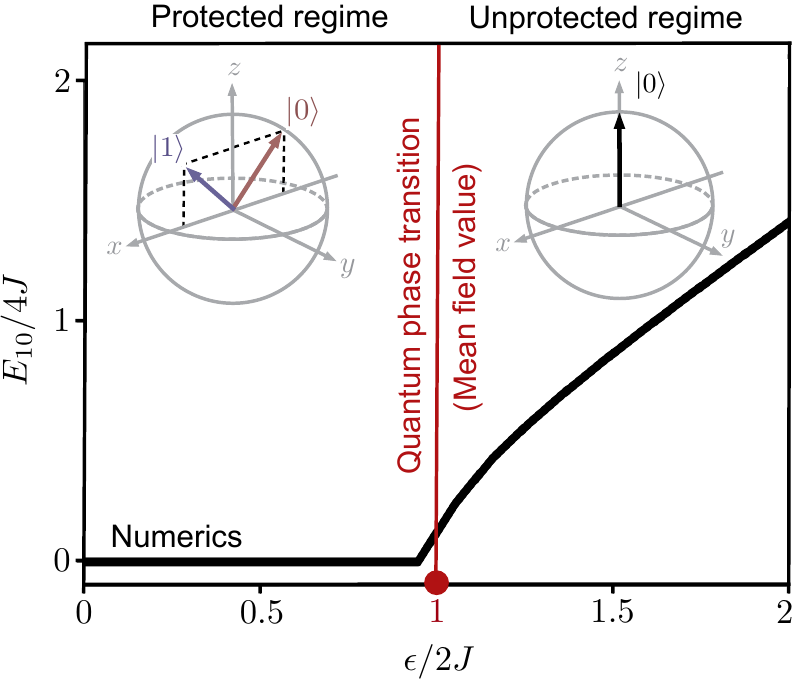}
    \caption{\textbf{`Giant spin' representation.} The effective Hamiltonian for an $N$-interferometer device can be written in terms of an $N$-spin representation (or `giant spin'). In the large-$N$ limit, two phases can be distinguished as a function of the flux noise amplitude $\epsilon$ and ferromagnetic exchange coupling $J$: In the protected regime for $\epsilon<2J$ (left panel), the ground state subspace $\{|0\rangle,|1\rangle\}$ is two-fold degenerate and the mean spin direction of the ground states is in the $xz$-plane with angles $\pm\theta_{0}\approx\arccos(\epsilon/2J)$ relative to the $z$-axis. In the unprotected regime for $\epsilon>2J$ (right panel) the ground state $\{|0\rangle\}$ is non-degenerate with a mean spin direction along the $z$-axis. The protected and unprotected regimes are separated by a quantum phase transition that, in a mean field picture, is located at $\epsilon/2J=1$ (vertical red line). From a numerical exact diagonalization, we can compute the energy splitting $E_{10}$ between the two lowest energy states as a function of $\epsilon/2J$ (solid black curve). We find that the mean field value of the quantum phase transition is approached from below by increasing the number of interferometers. The plot shows $E_{10}/4J$ for $N=2000$ interferometers.
    }
    \label{fig7}
\end{figure}

To find the variational parameters $\theta$ and $\chi$ of this mean-field ansatz, we 
minimize the variational energy,
\begin{equation}
\begin{split}
E^{(N)}_{\text{var}}(\theta,\chi)&\equiv
\left\langle \theta,\chi\right|
H^{(N)}_{\text{eff}}
\left|
\theta,\chi
\right\rangle
\\
&=
-\epsilon N \cos\theta
-
J(N-1)
(\sin\theta\,
\cos\chi)^{2}.
\end{split}
\end{equation}
From the minimization, we identify two phases with different ground state degeneracies:

The first phase occurs for the parameter range $|\epsilon|< 2J$. In this phase, the ground state subspace is doubly degenerate. The two spin coherent states with  $\theta_{0}\approx\arccos(\epsilon/2J)$ and $\chi_0=0,\pi$ form a basis of the subspace and realize the two qubit states. For vanishing flux noise, $\epsilon\rightarrow 0$, we note that the two qubit states correspond to the ferromagnetic configurations, $\{\left|\rightarrow,\dots,\rightarrow\right\rangle,\left|\leftarrow,\dots,\leftarrow\right\rangle\}$, that we already encountered for the $N=2$ loop device in the previous section. For finite flux noise, $\epsilon> 0$, the two qubit states acquire a small rotation angle in the $xz$ plane so that $\langle S_{z}\rangle=\epsilon/2J$, see the left panel of Fig.\,\ref{fig7}. However, despite the rotation, the two states remain degenerate as long as $|\epsilon|< 2J$, which sets an upper bound for the flux noise protection of our qubit in the limit $N\gg 1$.

The second phase occurs for the parameter ranges $|\epsilon|\geq 2J$. For our multi-loop device, this parameter range corresponds to the situation of substantial flux noise. As expected for such a scenario, the ground state is non-degenerate and spanned by the spin-coherent state with $\theta_{0}=0$, which points along the $z$-axis with $\langle S_{z}\rangle = 1$, see the right panel of Fig.\,\ref{fig7}. The encoding of a protected qubit is not possible in this regime. 

We further note that the `mean field' transition point that separates the two phases at $|\epsilon|=2J$ marks a quantum phase transition point, since the two-fold degenerate ground state subspace abruptly changes to a single non-degenerate ground state. From a numerical diagonalization of the Hamiltonian in Eq.\,\eqref{Eq20}, we have found that the quantum phase transition point of the `mean field' picture is approached from below upon increasing the length of the interferometer array, see Fig.\,\ref{fig7}. These findings are consistent with the results of the full model, for which we also found a broadening of the flux noise protection window upon increasing the length of the interferometer array, recall Fig.\,\ref{fig6}(c).

\section{Conclusions}
We have proposed a protected superconducting qubit realized in an array of superconducting islands connected by semiconductor Josephson interferometers. Such interferometers can realize $\cos2\phi$ elements when gate-tuned into balance and frustrated by a half-flux quantum \cite{larsen2020}. When an array of $\cos2\phi$ elements is shunted by a large capacitance, the qubit encoded in the degenerate ground state subspace is robust to offset charge and flux noise for a window around zero offset charge and half flux quantum. By introducing an effective spin model, we showed that flux noise protection broadens upon increasing the length of the array. In the long-array limit, a giant-spin model yielded a quantum phase transition as  function of flux offset between protected and unprotected regimes. The construction of interferometer array protected qubits can be realized using existing semiconductor-superconductor hybrid materials based on semiconductor nanowires \cite{larsen2020} or two-dimensional heterostructures \cite{casparis2018,oconnell2021,elfeky2021}.

\section*{Acknowledgements}
We thank Samuel Boutin, Reinhold Egger, Michael Freedman, Matthew Hastings, Morten Kjaergaard, and Patrick Lee for helpful discussions.  We acknowledge support from the Danish National Research Foundation, Microsoft, and a research grant (Project 43951) from VILLUM FONDEN.

\subsection*{Appendix A: Charging energies}
In this Appendix, we derive explicit expressions for the charging energies of the nanowire array qubit.
\\

To start, we write the Hamiltonian of the device in the following form, 
\begin{equation}
H
=
\frac{1}{2}
(\tilde{\boldsymbol{n}}-\tilde{\boldsymbol{n}}_{g})^{T}
\tilde{\boldsymbol{C}}^{-1}
(\tilde{\boldsymbol{n}}-\tilde{\boldsymbol{n}}_{g}) + V(\tilde{\boldsymbol{\phi}}),
\end{equation}
where $\tilde{\boldsymbol{n}}$ and $\tilde{\boldsymbol{n}}_{g}$ are vectors that contain the Cooper pair number operators and the offset charges of the $N+1$ superconducting islands, and $\tilde{\boldsymbol{\phi}}$ is the vector of phase operators. Moreover, $\tilde{\boldsymbol{C}}$ denotes the $(N+1)\times (N+1)$ capacitance matrix given by,
\begin{align}
&\tilde{\boldsymbol{C}} = 
\\
&\begin{bmatrix} 
    C_{B}+C_{S} & -C_{S} & 0  & & 0  & 0 & -C_{B} \\
    -C_{S} & 2C_{S} & -C_{S}  & \dots& 0  & 0 & 0 \\
    0 & -C_{S} & 2C_{S} &  & 0  & 0 & 0 \\
     & \vdots &  & \ddots &  & \vdots &    \\ 
    0 & 0 & 0 &  & 2C_{S} & -C_{S} & 0 \\
     0 & 0 & 0 &  \dots & -C_{S} & 2C_{S} & -C_{S} \\
    -C_{B} & 0 & 0 &   & 0 & -C_{S} & C_{B}+C_{S} \\
    \end{bmatrix}\nonumber.    
\end{align}
As a next step, we move from a description of charges on individual islands (`node charges') to a description of relative charges between neighboring islands (`branch charges'), as used in the main text. We achieve this change by the following transformations,
\begin{equation}
\begin{split}
&\boldsymbol{n}
=
(\boldsymbol{R}^{T})^{-1}
\tilde{\boldsymbol{n}}, 
\\
&\boldsymbol{\phi}
=
(\boldsymbol{R}^{T})^{-1}
\tilde{\boldsymbol{\phi}}, 
\\
&\boldsymbol{n}_{g}
=
(\boldsymbol{R}^{T})^{-1}
\tilde{\boldsymbol{n}}_{g},
\\
&\boldsymbol{C}^{-1}
\equiv
\boldsymbol{R}\cdot 
\tilde{\boldsymbol{C}}^{-1}\cdot
\boldsymbol{R}^{T},
\end{split}
\end{equation}
where we have introduced the $(N+1)\times (N+1)$ transformation matrix,
\begin{equation}
\begin{split}
\boldsymbol{R} = 
\begin{bmatrix} 
    -1 & 1 & 0  & & 0  & 0 & 0 \\
    0 & -1 & 1  & \dots& 0  & 0 & 0 \\
    0 & 0 & -1 &  & 0  & 0 & 0 \\
     & \vdots &  & \ddots &  & \vdots &    \\ 
    0 & 0 & 0 &  & -1 & 1 & 0 \\
     0 & 0 & 0 &  \dots & 0 & -1 &1 \\
    1 & 1 & 1 &   & 1 & 1 & 1 \\
    \end{bmatrix}.    
\end{split}
\end{equation}
After removing the free mode in the circuit \cite{ding2021}, the resulting transformed Hamiltonian takes on the form,
\begin{equation}
H
=
\frac{1}{2}
(\boldsymbol{n}-\boldsymbol{n}_{g})^{T}
\boldsymbol{C}^{-1}
(\boldsymbol{n}-\boldsymbol{n}_{g})+ V(\boldsymbol{\phi}),
\end{equation}
with the transformed capacitance matrix,
\begin{equation}
\begin{split}
&\boldsymbol{C} = 
\\
&\begin{bmatrix} 
    C_{B}+C_{S} & C_{B} &  \dots& C_{B} & C_{B} \\
   C_{B} & C_{B}+C_{S} &  & C_{B}  & C_{B}  \\
     \vdots&  & \ddots &  & \vdots &    \\ 
     C_{B} & C_{B} &     & C_{B}+C_{S} & C_{B} \\
    C_{B} & C_{B} & \dots    & C_{B} & C_{B}+C_{S} \\
    \end{bmatrix}.    
\end{split}
\end{equation}

\begin{figure}
    \centering
    \includegraphics[width = \columnwidth]{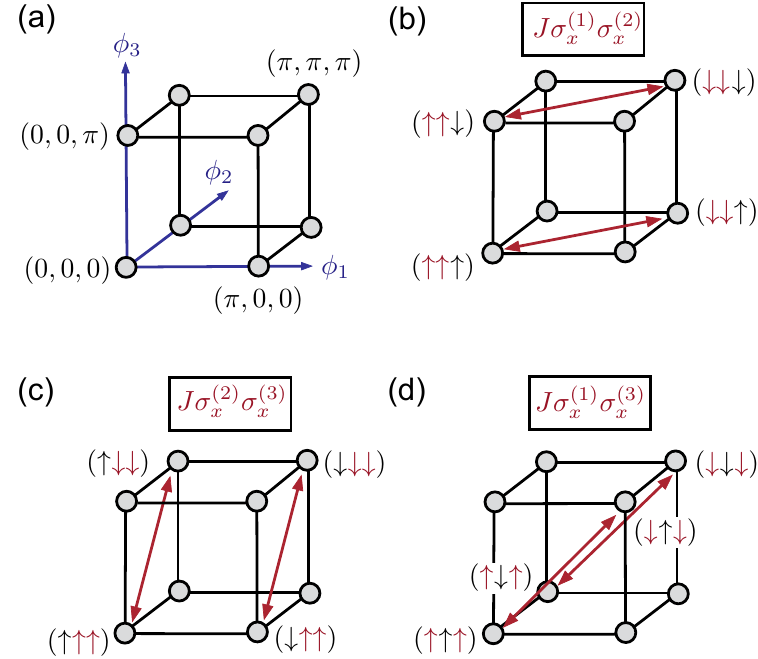}
    \caption{\textbf{Three-interferometer qubit} (a) The Josephson potential $V(\phi_{1},\phi_{2},\phi_{3})=-\sum_{i=1}^{3}E^{(i)}_{J,2}\cos(2\phi_i)$ of the three-interferometer qubit comprises $2^{3}=8$ minima (gray dots) that are located at $(\nu_{1}\pi,\nu_{2}\pi,\nu_{3}\pi)$ with $\nu_{1,2,3}=0,1$. (b) The tunneling (red arrows) between the minima in the $(\phi_{1},\phi_{2})$-plane induces a spin interactions $J\sigma^{(1)}_{x}\sigma^{(2)}_{x}$. (c) Same as (b) but for tunneling between the  minima in the $(\phi_{2},\phi_{3})$-plane, which induces a spin interaction $J\sigma^{(2)}_{x}\sigma^{(3)}_{x}$. (d) Same as (c) but for tunneling between the minima in the $(\phi_{1},\phi_{3})$-plane, which induces a spin interaction $J\sigma^{(1)}_{x}\sigma^{(3)}_{x}$.} 
    \label{fig8}
\end{figure}

To evaluate $\boldsymbol{C}^{-1}$ and obtain the relevant charging energies, we note that it can be written in the form,
\begin{equation}
\boldsymbol{C}
=
C_{B}\boldsymbol{Q}
+
C_{S}\boldsymbol{1},
\end{equation}
where we have defined the $N\times N$ matrices $(\boldsymbol{Q})_{ij}=1$ and $(\boldsymbol{1})_{ij}=\delta_{ij}$. We note that $\boldsymbol{Q}^{2}=N\boldsymbol{Q}$.
For the inverted capacitance matrix $\boldsymbol{C}^{-1}$, we make the ansatz,
\begin{equation}
\boldsymbol{C}^{-1}
=
\kappa
\boldsymbol{Q}
+
\frac{1}{C_{S}}\boldsymbol{1},
\end{equation}
where $\kappa$ is a to-be-determined parameter. We then require that,
\begin{equation}
\begin{split}
\boldsymbol{1} &\stackrel{!}{=}
\boldsymbol{C}\cdot
\boldsymbol{C}^{-1}\\
&=
\left[C_{B}\boldsymbol{Q}
+
C_{S}\boldsymbol{1}\right]
\left[
\kappa
\boldsymbol{Q}
+
\frac{1}{C_{S}}\boldsymbol{1}
\right]\\
&=
\left[
C_{B}\kappa N
+\frac{C_{B}}{C_{S}}
+C_{S}\kappa
\right]
\boldsymbol{Q}
+
\boldsymbol{1}.
\end{split}
\end{equation}
This condition is equivalent to,
\begin{equation}
C_{B}\kappa N
+\frac{C_{B}}{C_{S}}
+C_{S}\kappa
 \stackrel{!}{=}
 0,
\end{equation}
which leads us to,
\begin{equation}
 \kappa
=
-
\frac{C_{B}}{C_{S}(C_{S}+C_{B}N)}.   
\end{equation}
Having derived the explicit form of $\kappa$, we find that the inverse capacitance matrix of our setup is given by,
\begin{equation}
\boldsymbol{C}^{-1}
=
-
\frac{C_{B}}{C_{S}(C_{S}+C_{B}N)}
\boldsymbol{Q}
+
\frac{1}{C_{S}}\boldsymbol{1},
\end{equation}
or, equivalently, 
\begin{equation}
(\boldsymbol{C}^{-1})_{i,j}
=
\begin{cases} \frac{1}{C_{S}}-
\frac{C_{B}}{C_{S}(C_{S}+C_{B}N)} &\mbox{for } i = j \\ 
-
\frac{C_{B}}{C_{S}(C_{S}+C_{B}N)} & \mbox{otherwise}  \end{cases}.
\end{equation}
For the case of $N=2$ interferometers that was discussed in the main text, we find 
\begin{equation}
(\boldsymbol{C}^{-1})_{i,j}
=
\begin{cases} 
\frac{C_{B}+C_{S}}{2C_{B}C_{S}+C^{2}_{S}}
&\mbox{for } i = j \\ 
-\frac{C_{B}}{2C_{B}C_{S}+C^{2}_{S}}& \mbox{otherwise}  \end{cases}.
\end{equation}    
For the case of $N=3$ interferometers, we find 
\begin{equation}
(\boldsymbol{C}^{-1})_{i,j}
=
\begin{cases} 
\frac{2C_{B}+C_{S}}{3C_{B}C_{S}+C^{2}_{S}}
&\mbox{for } i = j \\ 
-\frac{C_{B}}{3C_{B}C_{S}+C^{2}_{S}}& \mbox{otherwise}  \end{cases}.
\end{equation}    
It is important to note that the inverse capacitance matrix not only includes nearest-neighbor elements, $(\boldsymbol{C}^{-1})_{i,i+1}$, but also beyond-nearest-neighbor elements, $(\boldsymbol{C}^{-1})_{i,j}$ with $j\neq i+1$. For the example of $N=3$ interference loops shown in Fig.\,\ref{fig8}, this implies that the total effective spin interaction not only includes the nearest-neighbor spin interactions, $J\sigma^{(1)}_{x}\sigma^{(2)}_{x}$ and $J\sigma^{(2)}_{x}\sigma^{(3)}_{x}$, but also the beyond-nearest-neighbor spin interaction, $J\sigma^{(1)}_{x}\sigma^{(3)}_{x}$.

\subsection*{Appendix B: Tight-binding models}
Now, we provide more details on the derivation of the tight-binding models for the single- and two-interferometer qubits.

We begin with the circuit Hamiltonian of the single-interferometer qubit,
\begin{equation}
    H=4E_C n^2 - E_{J,2}\cos2\phi.
\end{equation}
We recall that the eigenfunctions are Bloch states $\psi_{\ell,n_g}(\phi)$, which obey the quasi-periodic boundary conditions,
\begin{equation}
\label{Eq39}
\psi_{\ell,n_g}(\phi+2\pi)=e^{in_g2\pi}\psi_{\ell,n_g}(\phi),
\end{equation}
where $\ell$ is the band-index.

For the derivation of the tight-binding model of the single-interferometer qubit, we consider the eigenfunctions of the two lowest-energy states, $\psi_{0,n_g}(\phi)$ and $\psi_{1,n_g}(\phi)$, and combine them as follows, 
\begin{equation}
\begin{split}
\psi_{\uparrow,n_g}(\phi)=\frac{1}{\sqrt{2}}
\left[
\psi_{0,n_g}(\phi)-\psi_{1,n_g}(\phi)
\right],
\\
\psi_{\downarrow,n_g}(\phi)=\frac{1}{\sqrt{2}}
\left[
\psi_{0,n_g}(\phi)+\psi_{1,n_g}(\phi)
\right].
\end{split}
\end{equation}
These combinations of the eigenfunctions have finite support in the 0 or $\pi$ valleys of the $\cos2\phi$ potential. We associate these Bloch states with the spin-up and spin-down states of the effective spin model, and introduce the following Dirac notation $\langle\phi\left|\downarrow\right\rangle = \psi_{\downarrow,n_g}(\phi)$ and $\langle\phi\left|\uparrow\right\rangle=\psi_{\uparrow,n_g}(\phi)$. 

We can use the Wannier functions to express the Bloch states, 
\begin{equation}
\begin{split}
\left|\downarrow\right\rangle&=
\frac{1}{\sqrt{M}}
\sum_{m}
e^{in_g2\pi m}
\left|w_{\downarrow,m}\right\rangle,
\\
\left|\uparrow\right\rangle&=
\frac{1}{\sqrt{M}}
\sum_{m}
e^{in_g2\pi m}
\left|w_{\uparrow,m}\right\rangle,
\end{split}
\end{equation}
where $M$ is a normalization factor, and for simplicity we again introduced the Dirac notation for the Wannier states $\langle\phi\left|w_{\downarrow,m}\right\rangle = w_\downarrow(\phi-2\pi m)$ and $\langle\phi\left|w_{\uparrow,m}\right\rangle = w_\uparrow(\phi-2\pi m)$.

We are now in the position to write down the effective tight-binding Hamiltonian by projecting the Hamiltonian onto the subspace spanned by the spin states $\{\left|\uparrow\right\rangle,\left|\downarrow\right\rangle\}$, which yields,
\begin{equation}
\label{Eq42}
H_{\text{eff}}
=
\sum_{s,s'=\uparrow,\downarrow}
\langle s|H|s'\rangle\,
|s\left\rangle\right\langle s'|.
\end{equation}

\begin{figure}
    \centering
    \includegraphics[width = \columnwidth]{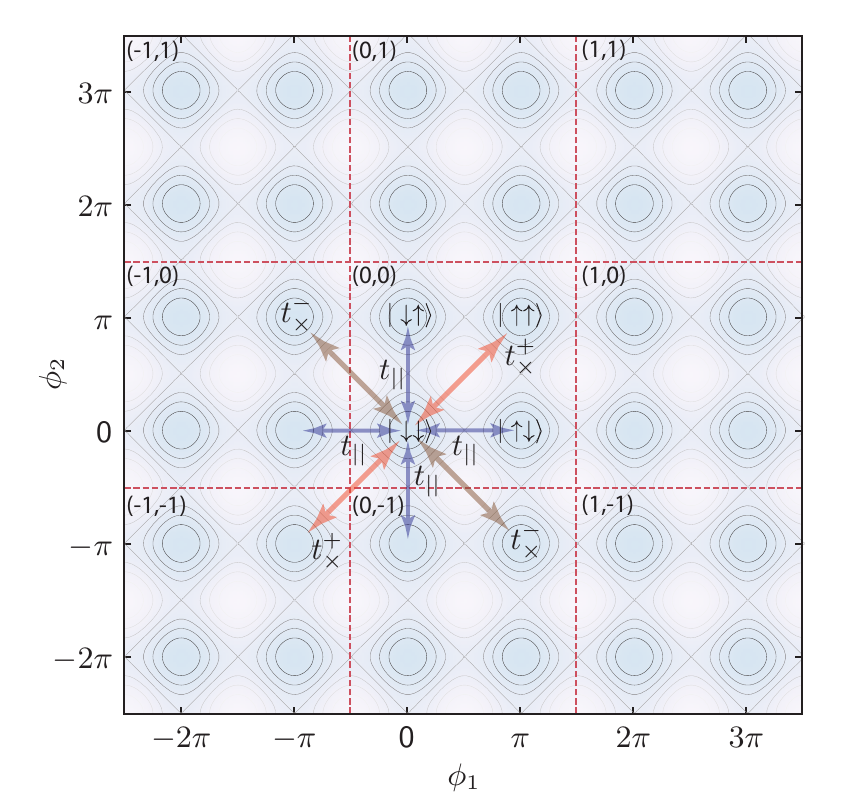}
    \caption{\textbf{Tight-binding model on the two-dimensional phase lattice} The extend $\cos2\phi$ potential of the $N=2$ array. Red dashed lines display the boundary of the unit cells. The unit cell indices $(l,m)$ are shown in the upper left corners of the unit cells. The location of the spin states in the (0,0) cell are indicated by $|s_1s_2\rangle$. To find the effective spin model, we need to take into account many possible hybridization directions. As an example, the arrows show the hybridizations involving the $\left|\downarrow\downarrow\right\rangle$ state.} 
    \label{fig9}
\end{figure}

Next, we can compute the effective Hamiltonian by evaluating the following matrix elements in a tight-binding approximation ($E_{J,2}\gg E_{C}$), 
\begin{align}
\label{Eq43}
\left\langle \uparrow\right|H\left|\downarrow\right\rangle 
&=
\frac{1}{M}
\sum_{m,m'}
e^{in_g2\pi (m-m')}
\left\langle w_{\uparrow,m'}\right|H\left|w_{\downarrow,m}\right\rangle\nonumber \\
&= 
\frac{1}{M}
\sum_{m}
e^{-in_g2\pi m}
\left\langle w_{\uparrow,m}\right|H\left|w_{\downarrow,0}\right\rangle 
\\
&\approx
e^{i2\pi n_g}
\left\langle w_{\uparrow,-1}\right|H\left|w_{\downarrow,0}\right\rangle 
+
\left\langle w_{\uparrow,0}\right|H\left|w_{\downarrow,0}\right\rangle\nonumber.
\end{align}

We also note that $\left\langle \uparrow\right|H\left|\uparrow\right\rangle =\left\langle \downarrow\right|H\left|\downarrow\right\rangle$, so that the diagonal terms in the effective Hamiltonian only give a constant offset.

\begin{figure}
    \centering
    \includegraphics[width = \columnwidth]{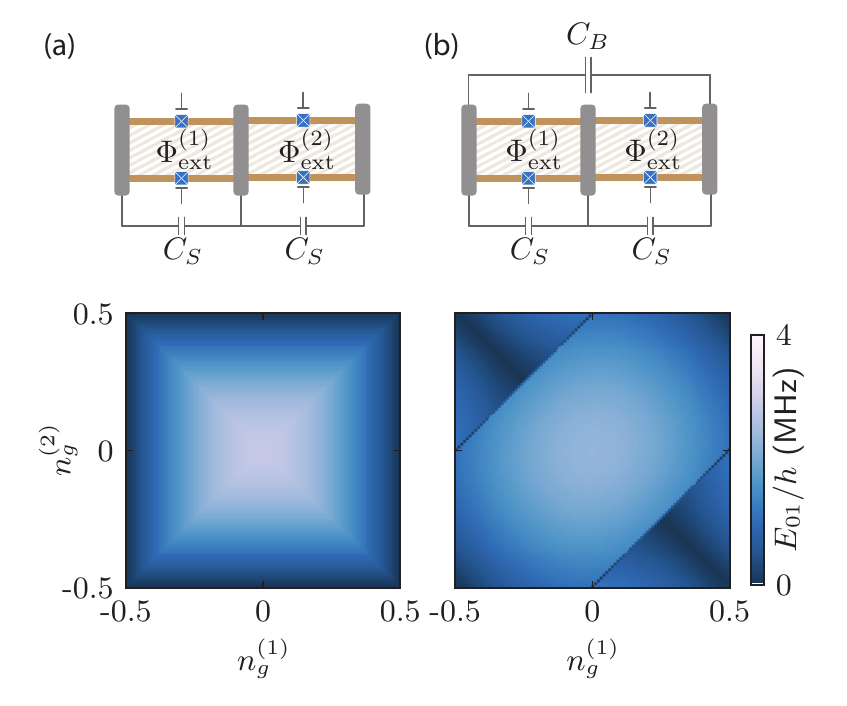}
    \caption{\textbf{Energy dispersion as function of offset charges} (a) The qubit energy in the unprotected regime shows a maximum and a sweet spot at $n_g^{(1)}=n_g^{(2)}=0$, and gradually decreases as the offset charges are increased. (b) The offset-charge dependence in the protected regime displays a more complicated pattern due to the offset charge dependence of the coefficient of the $\sigma_x^{(1)}\sigma_x^{(2)}$ interaction. The qubit can be operated around $n_g^{(1)}=n_g^{(2)}=0$ sweet spot. Parameters are the same as in Fig.\,\ref{fig6}.} 
    \label{fig10}
\end{figure}

It is now helpful to introduce the inter- and intra-unit cell tunnelling amplitudes,
\begin{equation}
\begin{split}
t_{\text{in}}&=   \left\langle w_{\uparrow,0}\right|H\left|w_{\downarrow,0}\right\rangle
 \\
  t_{\text{out}}&=   \left\langle w_{\uparrow,-1}\right|H\left|w_{\downarrow,0}\right\rangle.
\end{split}
\end{equation}
By noting that $t_{\text{in}}=t_{\text{out}}\equiv t$, we can write the effectiv Hamiltonian of Eq.\,\eqref{Eq42} compactly as, 
\begin{equation}
\begin{split}
H_{\text{eff}}&=
t(1+e^{i2\pi n_g})|\uparrow\left\rangle\right\langle \downarrow|+\text{H.c.}
\\
&=
\begin{bmatrix}
0 & t(1+e^{i2\pi n_g}) \\
t(1+e^{-i2\pi n_g}) & 0 
\end{bmatrix}.
\end{split}
\end{equation}
After introducing the spin-space Pauli matrices $\sigma_{x,y,z}$ as in the main text, we find that this form of effective Hamiltonian reads as, 
\begin{equation}
H_{\text{eff}}  =2t\cos(\pi n_g)\left\{\cos(\pi n_g)\sigma^x + \sin(\pi n_g)\sigma^y\right\},
\end{equation}
which is the result that we presented in Eq.\,\eqref{Eq6}.

Now, we follow a similar approach to obtain the spin model for two coupled interferometers. Using Dirac notation, we express the two-dimensional Bloch functions, $|s_1s_2\rangle$ with the two-dimensional Wannier functions, $|w_{s_1, s_2,\ell,m}\rangle$
\begin{equation}
    \left|s_1s_2\right\rangle=
\frac{1}{M}
\sum_{\ell,m}
e^{in_g^{(1)}2\pi \ell}e^{in_g^{(2)}2\pi m}
\left|w_{s_1, s_2,\ell,m}\right\rangle
\end{equation}

To get the effective spin Hamiltonian, we first project the Hamiltonian onto the subspace of the four lowest lying Bloch states, $\{\left|\downarrow\downarrow\right\rangle,\left|\downarrow\uparrow\right\rangle,\left|\uparrow\downarrow\right\rangle,\left|\uparrow\uparrow\right\rangle\}$ 
\begin{equation}
\label{Eq42}
H_{\text{eff}}^{(2)}
=
\sum_{\substack{s_1,s_2,\\s_1',s_2'=\uparrow,\downarrow}}
\langle s_1s_2|H|s_1's_2'\rangle\,
|s_1s_2\left\rangle\right\langle s_1's_2'|.
\end{equation}

When the potential is symmetric (the loops are biased at half flux quantum and the junctions are balanced), there are three different types of hybridization: parallel to the $\phi_i$ axes, $t_{||}$, along the $\phi_1+\phi_2$ direction,  $t_{\times}^+$, and along the $\phi_1-\phi_2$ direction, $t_{\times}^-$, see Fig.\,\ref{fig9}. In the protected regime, the wavefunctions are elongated along the $\phi_1-\phi_2$ direction, thus, $t_{\times}^+\ll t_{\times}^-$. After taking into account all possible terms, using the tight-binding approximation, and expressing the result with Pauli matrices, we arrive at the effective Hamiltonian of two coupled interferometers,
\begin{equation}
\begin{split}
H_{\text{eff}}^{(2)} &= 
2t_{||}\left[\cos(\pi n_g^{(1)})\tilde\sigma_{x}^{(1)}+\cos(\pi n_g^{(2)})\tilde\sigma_{x}^{(2)}\right] + \\ 
& 2t_{\times}^+\cos\left(\pi \left[n_g^{(1)}+n_g^{(2)}\right]\right)\tilde\sigma_{x}^{(1)}\tilde\sigma_{x}^{(2)} + \\
&2t_{\times}^-\cos\left(\pi \left[n_g^{(1)}-n_g^{(2)}\right]\right)\tilde\sigma_{x}^{(1)}\tilde\sigma_{x}^{(2)},
\end{split}
\end{equation}
which is equivalent to the result in Eq.\,\eqref{Eq18}, after taking into account that $t_{\times}^+\ll t_{\times}^-$, and $J=-t_{\times}^-$.

\subsection*{Appendix C: Offset-charge sensitivity}

In Fig.\,\ref{fig10}, we provide numerical results on the charge sensitivity of the unprotected and protected $N=2$ array. In the protected regime, the energy is a complicated function of the offset charges because of the offset-charge dependence of the interaction term, $2t_{\times}^-\cos\left(\pi \left[n_g^{(1)}-n_g^{(2)}\right]\right)\tilde\sigma_{x}^{(1)}\tilde\sigma_{x}^{(2)}$. The qubit needs to be biased around the $n_g^{(1)}=n_g^{(2)}=0$ regime.

\end{document}